\documentclass{ws-ijmpe}

\begin{document}

\markboth{S. K. Biswal et al.}{Shell closure in the superheavy valley}

\catchline{}{}{}{}{}

\title{Search of double shell closure in the superheavy nuclei using
a simple effective interaction\\
}

\author{\footnotesize S. K. Biswal, M. Bhuyan, S. K. Singh 
and S. K. Patra 
}

\address{Institute of Physics, Sachivalaya Marg, 
Bhubaneswar-751 005, India. \\
}

\maketitle

\begin{history}
\received{(received date)}
\revised{(revised date)}
\end{history}

\begin{abstract}
This paper refers to an another attempt to search for spherical double
shell closure nuclei beyond {\it Z}=82, {\it N}=126. All calculations
and results are based on a newly developed approach entitled as simple
effective interaction. Our results predict the combination of magic
nucleus occurs at {\it N}=182 ({\it Z}=114,120,126). All possible evidences
for the occurrence of magic nuclei are discussed systematically. And, the
obtained results for all observables compared with the relativistic mean
field theory for NL3 parameter.
\end{abstract}

\section{Introduction}
Starting from the discovery of nucleus, the formation of new element is
an interesting topic in Nuclear Physics. So far the synthesis of heaviest
element in laboratory is {\it Z}=118 in the hot fusion reaction process at
JINR Dubna \cite{ogan06,ogan12}. The possibility of the existence (synthesis)
of these superheavy elements is mainly due to the attractive shell corrections
against the destructive Coulomb repulsion. Although atomic number Z=114 was
predicted to be the next magic number after {\it Z}=82 and neutron number
{\it N}=184, recently attention has shifted to the nucleus {\it Z}=120
with {\it N}=182/184 \cite{patra97,rutz97,bhuyan12,sil04}. The experimental
discovery  of the superheavy elements also support this prediction to some
extent. Thus, the synthesis of {\it Z}=120 is in full swing at the worlds'
most laboratories like, Dubna (Russia), RIKEN (Japan), GSI (Germany).

Using cold fusion reaction, elements from $Z=107-112$ are synthesized
at GSI \cite{hofman00,hofman81,hofman1995,hofman95,hofman96,hofman98,hofman09}.
At the production time of {\it Z} = 112 nucleus at GSI, the fusion cross
section was extremely small ($1$ pb) \cite{hofman96}, which led to the
conclusion that reaching still heavier elements will be very difficult
by this process. The element {\it Z}=113 was also synthesized in cold-fusion
reaction at RIKEN with a very low cross section $\sim 0.03$ pb \cite{morita07}
confirming the limitation of cold-fusion synthesis. To overcome this problem
in hot fusion evaporation reaction with  deformed actinide targets and
 neutron-rich doubly magic spherical projectile like $^{48}Ca$ are used
in the production of superheavy nuclei $Z = 112 - 118$ at Dubna
\cite{ogni98,ogni01,ogni04,ogni07,ogni10,ogni11}.

It is thus a matter of challenge for every theoretical prediction in
nuclear  physics to find suitable combination of proton and neutron,
which gives double closure shell nuclei beyond $^{208}$Pb and will be
the next element of epicenter for experimental synthesis. In the present
work, our aim is to look for a suitable combination of proton and neutron
in such a way that the resultant combination will be the next magic nucleus
after $^{208}$Pb. This work is not a new, but a revisit
of our earlier prediction with in a new simple effective interaction (SEI).
The SEI interaction is recently developed by us \cite{bhuyan13} and
given a parameter set which is consistent with both nuclear matter and
finite nuclei. Here, we have used this
SEI interaction. A systematic investigation of
the nuclear structure is done and reconfirmed the double closed nucleus
as Z=120 with N=182/184.

The paper is organized as follows: In Sec. II, the theoretical formalism
of the SEI is presented. The procedures for numerical calculations to estimate
the binding energy and root mean square radii are outlined. The results
and discussions are given in Sec. III. The characteristics of magic structure
of nucleus using two neutron separation energy, pairing gap of proton and
neutron are analyzed for superheavy region. In this section stability of
such nuclei are also studied in terms of the chemical potentials. Finally
a summary and a concluding remarks are given in Sec. IV.

\section{The Theoretical Framework}
\subsection{Simple Effective Interaction}

The present formalism is based on a simple way to make a consistent
parametrization for both finite nucleus and infinite nuclear matter
with a momentum dependence finite range term of conventional form,
such as {\it Yukawa}, {\it Gaussian} or {\it exponential} to the
standard Skyrme interaction \cite{trr98,trr02,bhuyan13}. We have used
the technique of Refs. \cite{trr98,trr02,bhuyan13} considering a Gaussian
term as the momentum dependence finite range interaction simulation
the effect of Gogny type interaction \cite{gogny80,gogny84}. Then it is
applied to nuclear equation of state as well as to finite nuclei through
out the periodic table \cite{bhuyan13}. The Hartree-Fock (HF) formalism is
adopted to calculate the wave-function of the nuclear system which then used
to evaluate the nuclear observables, such as binding energy, root mean square
radii etc. The detail formalism and numerical procedure can be found in
\cite{bhuyan13}. The form of the simple effective interaction (SEI) is given
by \cite{bhuyan13}:
\begin{eqnarray}
v_{eff}({\bf r})&=&t_0 (1+x_0P_{\sigma})\delta({\bf r}) \nonumber \\ 
&&+t_3(1+x_3 P_{\sigma})\left(\frac{\rho({\bf R})}{1+b\rho({\bf R})}\right)^{\gamma} 
\delta({\bf r}) \nonumber \\
&&+ \left(W+BP_{\sigma}-HP_{\tau}-MP_{\sigma}P_{\tau}\right)f({\bf r}) \nonumber \\
&&+i W_{0}({\bf \sigma}_i+{\bf \sigma}_j)({\bf k'}\times{\delta}({\bf r}_i+{\bf r}_j){\bf k}).
\end{eqnarray}
Where, $f({\bf r})$ is the functional form of the finite
range interaction containing a single range parameter $\alpha$. The
finite range Gaussian form is given as $f({\bf r})=e^{-r^2/\alpha^2}$. The
other terms having their usual meaning \cite{bhuyan13}.
To prevent the supra luminous behaviour of the nuclear matter, the
usual value of $b$ \cite{trr05,trr07,trr09,trr11} is taken.
There are $11$-parameters in the interaction, namely
$t_0$, $x_0$, $t_3$, $x_3$, $b$, $W$, $B$, $H$, $M$, $\gamma$ and $\alpha$.
The expression for energy density, single particle energy and other
relevant quantities are obtained from Eqn. (1) for Gaussian $f(r)$ 
defined in Ref. \cite{bhuyan13}. The numerical values of the parameter
set, SEI and RMF(NL3) are given in Table I.

\begin{table}
\caption{\label {tab:table 2}{The value of interaction parameters 
for simple effective interaction (SEI) and RMF (NL3) [33] sets and 
their nuclear matter properties at saturation.}}
\renewcommand{\tabcolsep}{0.6cm}
\renewcommand{\arraystretch}{1.0}
\begin{tabular}{|cc|cc|cccccccc}
\hline
\multicolumn{2}{|c|}{SEI} & \multicolumn{2}{c|}{RMF (NL3)} \\
\hline
$\gamma$                    & $\frac{1}{2}$ & ${M} $ (MeV)         & 939 \\
$b$ ($fm^3$)                & 0.5914        & ${m}_{\sigma} $ (MeV)& 508.1941 \\
$t_0$ ($MeVfm^3$)           & 437.0         & ${m}_{\omega} $ (MeV)& 782.6010 \\
$x_0$                       & 0.6           & ${m}_{\rho}   $(MeV) & 7630.0 \\
$t_3$ ($MeVfm^{3(\gamma+1)})$ & 9955.2      & ${g}_{\sigma} $      & 10.2169 \\
$x_3$                       & -0.1180       & ${g}_{\omega} $      & 12.8675 \\
W (MeV)                     & -589.09       & ${g}_{\rho}  $       & 8.9488 \\
B (MeV)                     & 130.36        & ${g}_2$  ($fm^{-1}$) & -10.4307 \\
H (MeV)                     & -272.42       & ${g}_3$              & 28.8851 \\
M (MeV)                     & -192.16       &                      &          \\
$\alpha$ (fm)               &  0.7596       &                      &          \\ 
$W_0$ (MeV)                 & 115.0         &                      &           \\
\hline
\multicolumn{4}{|c|}{Nuclear matter}                                           \\
\hline
$\rho_o$ ($fm^{-3}$)       & 0.157           & $\rho_o$ ($fm^{-3}$)& 0.148 \\
$e(\rho_0)$ (MeV)          & -16.0           & $e(\rho_0)$ (MeV)   & -16.24  \\
$E_{s}$ (MeV)              & 35.0            & $E_{s}$ (MeV)       & 37.4  \\
$K_0$ (MeV)                & 245             & $K_0$ (MeV)         & 271.5  \\
\hline
\hline
\end{tabular}
\end{table}

\subsection{Relativistic mean field (RMF) formalism}

The starting point of the RMF theory is the basic Lagrangian containing
nucleons interacting with $\sigma-$, $\omega-$ and $\rho-$meson fields.
The photon field $A_{\mu}$ is included to take care of the Coulomb
interaction of protons. The relativistic mean field Lagrangian density
is expressed as~\cite{boguta97,walecka},
\begin{eqnarray}
L & = & \overline{\psi_{i}}\{i\gamma^{\mu}\partial_{\mu}-M\}\psi_{i}+{\frac12}\partial^{\mu}
\sigma\partial_{\mu}\sigma-{\frac12}m_{\sigma}^{2}\sigma^{2}\nonumber \\
&& -g_{\sigma}\overline{\psi_{i}}\psi_{i}
\sigma-{\frac14}\Omega^{\mu\nu}\Omega_{\mu\nu}
+{\frac12}m_{w}^{2}V^{\mu} V_{\mu}\nonumber \\
&& -g_{w}\overline\psi_{i}\gamma^{\mu} \psi_{i}V_{\mu}-{\frac14}
\vec{B}^{\mu\nu}.\vec{B}_{\mu\nu}
+{\frac12}m_{\rho}^{2}{\vec R^{\mu}} .{\vec{R}_{\mu}}\nonumber \\
&&-g_{\rho}\overline\psi_{i}\gamma^{\mu}\vec{\tau}\psi_{i}.\vec{R^{\mu}}-{\frac12}m_{\delta}^{2}
\delta^{2}+g_{\delta}\overline\psi_{i}\delta\vec{\tau}\psi_{i}.
\end{eqnarray}
Here M, $m_{\sigma}$, $m_{\omega}$ and $m_{\rho}$ are the masses for nucleon,
${\sigma}$-, ${\omega}$- and ${\rho}$-mesons and ${\psi}$ is the Dirac spinor.
The field for the ${\sigma}$-meson is denoted by ${\sigma}$, ${\omega}$-meson
by $V_{\mu}$ and ${\rho}$-meson by $R_{\mu}$.
$g_s$, $g_{\omega}$, $g_{\rho}$ and $e^2/4{\pi}$=1/137 are the coupling
constants for the ${\sigma}$, ${\omega}$, ${\rho}$-mesons and photon
respectively. $g_2$ and $g_3$ are the self-interaction coupling constants
for ${\sigma}$ mesons. By using the classical variational principle we
obtain the field equations for the nucleons and mesons.
A static solution is obtained from the equations of motion to describe
the ground state properties of nuclei.
The set of nonlinear coupled equations are solved self-consistently in
one dimensional coordinate \cite{patra01}.
The total energy of the system is given by
\begin{equation}
 E_{total} = E_{part}+E_{\sigma}+E_{\omega}+E_{\rho}+E_{c}+E_{pair}+E_{c.m.},
\end{equation}
where $E_{part}$ is the sum of the single particle energies of the nucleons and
$E_{\sigma}$, $E_{\omega}$, $E_{\rho}$, $E_{c}$, $E_{pair}$, $E_{cm}$ are
the contributions of the meson fields, the Coulomb field, pairing energy
and the center-of-mass motion correction energy, respectively.
We have used the well known NL3 parameter set~\cite{lala97} in our
calculations for RMF formalism.

\subsection{Pairing Correlation}

To take care of the pairing correlation for open shell nuclei the
constant gap, BCS-approach is used in our calculations.
The pairing energy expression is written as
\begin{equation}
E_{pair}=-G\left[\sum_{i>0}u_{i}v_{i}\right]^2,
\end{equation}
with $G$ is pairing force constant. The quantities $v_i^2$ and $u_i^2=1-v_i^2$ are the occupation
probabilities \cite{patra93,sero861,pres82}. The variational approach with
respect to $v_i^2$ gives the BCS equation
\begin{equation}
2\epsilon_iu_iv_i-\triangle(u_i^2-v_i^2)=0,
\end{equation}
using $\triangle=G\sum_{i>0}u_{i}v_{i}$.
The occupation number is defined as
\begin{equation}
n_i=v_i^2=\frac{1}{2}\left[1-\frac{\epsilon_i-\lambda}{\sqrt{(\epsilon_i-\lambda)^2+\triangle^2}}\right].
\end{equation}

The chemical potentials $\lambda_n$ and $\lambda_p$ are determined
by the particle number for protons and neutrons. The pairing energy
is computed as $E_{pair}=-\triangle\sum_{i>0}u_{i}v_{i}$. For a particular
value of $\triangle$ and $G$, the pairing energy $E_{pair}$ diverges if it
is extended to an infinite configuration space. In fact, in all realistic
calculations with finite range forces, $\triangle$ decreases with state for
large momenta near the Fermi surface. In the present case, we assume equal
pairing gap for all states $\mid\alpha>=\mid nljm>$ near the Fermi surface.
We use a pairing window, where the equations are extended up to the level
$\epsilon_i-\lambda\leq 2(41A^{1/3})$ \cite{bhu09}. The factor 2 has been 
determined so as to reproduce the pairing correlation energy for neutrons 
in $^{118}$Sn using Gogny force \cite{dech80,patra93,sero861}.

\section{Results and Discussions}
The quasi local Density Functional Theory (DFT) is used in this work, which is similar to
the one used by Hoffman and Lenske in Ref. \cite{hoffman98}.
The total energy is nothing but the sum of the
energy density contribution from different components of the
interaction along with spin-orbit and Coulomb term.
The energy density ${\cal H}_0$ for SEI set can be expressed as
\begin{eqnarray}
{\mathcal H}_0&=&\frac{\hbar^2}{2m}\left(\tau_n+\tau_p\right)+
{\mathcal H}_{d}^{Nucl}+{\mathcal H}_{exch}^{Nucl}+ {\mathcal H}^{SO} 
+{\mathcal H}^{Coul}.
\end{eqnarray}
>From this effective Hamiltonian $\tilde{H}$
we obtain the quasi local energy functional as:
\begin{eqnarray}
\varepsilon_0\left[\rho^{QL}\right]=\int{\mathcal H}_0d^3R.
\label{eq12}
\end{eqnarray}
The equations solved self-consistently to get the solution for
nucleonic system. Here we have taken only spherical solution for both RMF and SEI.

\begin{table}
\caption{\label{tab:table1}{The binding energy (BE) obtained from SEI
calculation is compared with the RMF(NL3) [33], finite range droplet 
model (FRDM) [39] and with experimental data of some of the known 
superheavy nuclei. The BE is in MeV.}}
\renewcommand{\tabcolsep}{0.6cm}
\renewcommand{\arraystretch}{1.0}
\begin{tabular}{|c|c|c|c|c|c|c|c|c|}
\hline
Element &\multicolumn{4}{c|}{BE}  \\
\hline
&SEI& RMF(NL3) &FRDM & Expt. \\
\hline
$^{258}{Md}$&1896.19&1897.70&1911.53&1911.69 \\
$^{258}{Rf}$&1884.95&1890.86&1905.25 &1904.69 \\
$^{261}{Rf}$&1906.38&1911.04&1924.28& 1923.93 \\
$^{259}{Db}$& 1886.94&1894.58 & 1907.00&1906.33  \\
$^{260}{Db}$& 1894.31& 1901.4 & 1913.34 & 1912.82\# \\
$^{260}{Sg}$& 1888.62& 1897.9 & 1909.90&1909.07  \\
$^{261}{Sg}$&1896.17&1905.02&1916.27& 1915.68 \\
$^{264}{Hs}$& 1906.86&1915.5 & 1927.62&1926.77 \\
$^{265}{Hs}$&1914.59&1922.9 &1934.40 &1933.50  \\
$^{269}{Ds}$&1932.81&1941.21&1952.06&1950.290\\
$^{285}{Fl}$&2029.41&2039.19&2044.12& 2040.03\# \\
$^{286}{Fl}$&2036.74&2046.17&2051.59& 2047.474\# \\
$^{287}{Fl}$&2043.36&2052.50&2057.65& 2053.19\# \\
$^{288}{Fl}$&2050.14&2058.73&2065.01& 2060.64\# \\
$^{289}{Fl}$&2056.80&2064.87&2071.04& 2066.06\#  \\
\hline
\hline
\end{tabular}
\end{table}

\subsection{Ground state binding energy}

The main objective of the present study is to find the double
shell closure in the superheavy valley. In this context, we
have concentrated on few observables such as separation energy $S_{2n}$,
chemical potential $\mu_n$, single-particle levels ${\cal E}_{n,p}$
and pairing energy $E_{pair}$. Before going to this unknown region
(superheavy valley), it is important to test our model for known
magic nuclei, which are experimentally and theoretically well
established. We calculate the binding energy of few known superheavy
nuclei using SEI. The obtained results are compared with RMF,
finite-Range-Droplet-Model (FRDM) \cite{moll97} and experimental
data \cite{audi12} in Table II. The \# marks in the experimental
column are for the extrapolated data from Ref. \cite{audi12}.
From the table, we find that the SEI and NL3 results are 
overestimated to the experimental values. A close observation of
the table shows the superiority of FRDM over SEI or NL3 for
lighter mass of the superheavy nuclei. In contrast to the lighter
region, the SEI predicts better results
for heavier isotopes. For example,  binding energy of $^{289}{Fl}$
is 2056.80 $MeV$ in SEI, whereas the values are 2064.87, 2071.04
and 2066.06 $MeV$ in RMF(NL3), FRDM and experiment (or systematics),
respectively. Based on this trend, one can expect that the prediction
of SEI gives us better insight about the magic structures
of superheavy nuclei in heavier mass region, which is the main objective
of the present investigation.

\begin{figure}
\resizebox{1.0\textwidth}{!}{\includegraphics{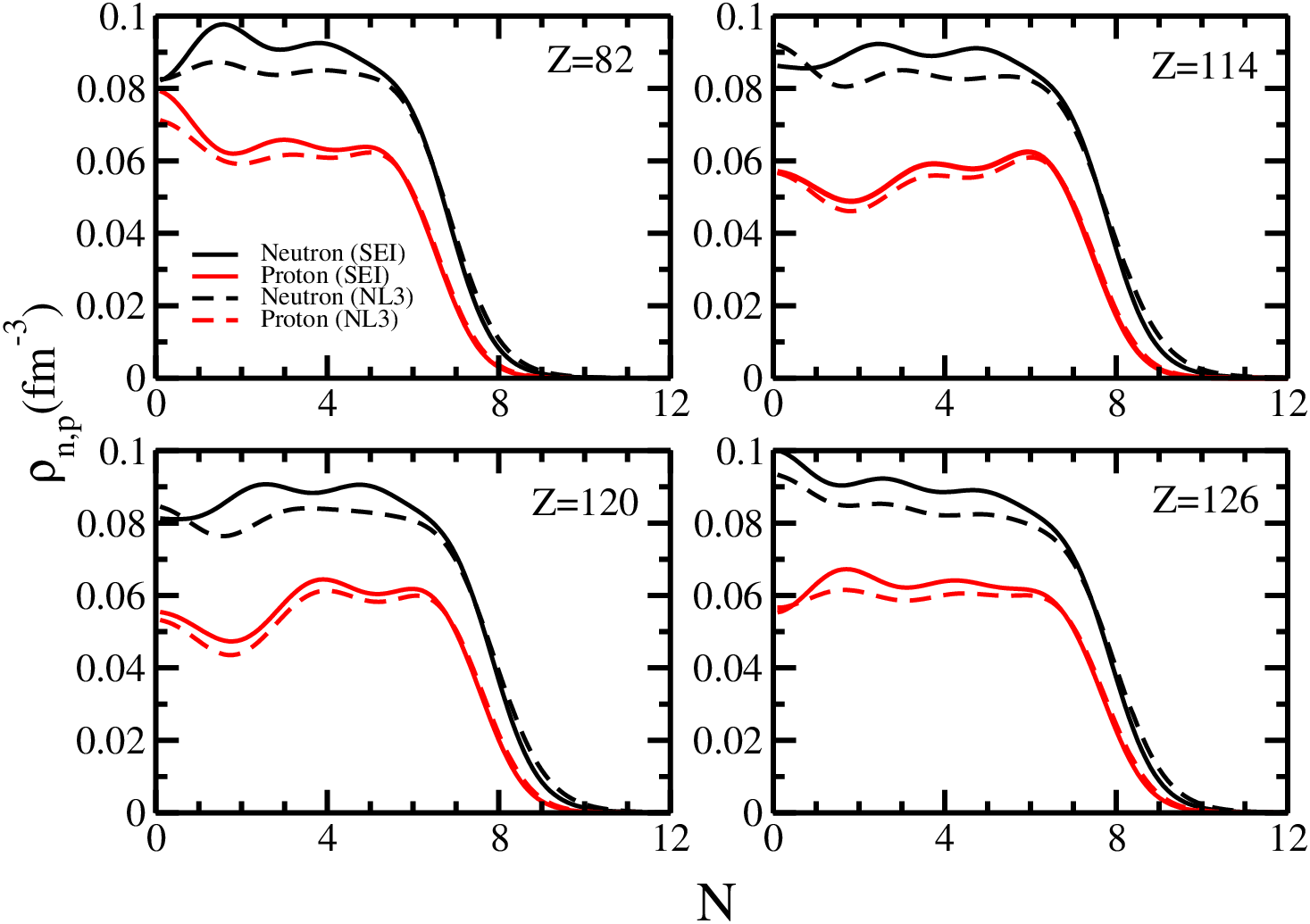}}
\caption{The ground state densities with SEI
for $^{208}$Pb, $^{298}$114, $^{304}$120 and $^{310}$126 are compared
with the RMF(NL3) results.}
\label{fig:1}
\end{figure}

\subsection{Density distribution of neutrons and protons}

After convinced with the binding energy of the superheavy nuclei,
we present the density distribution of protons and neutrons in
Figure 1. The densities are compared with the RMF(NL3) calculations.
In general, the RMF and SEI densities are  almost similar with
each other. However, a proper inspection reveals that the SEI
densities slightly over estimate the RMF(NL3) densities. This
overestimation is mostly at the middle region of the nucleus.
The humps at the central region for both the densities show shell
effect for all nuclei shown in the figure.

\subsection{Two neutron separation energy and location of closed shell}
\begin{figure}
\resizebox{1.0\textwidth}{!}{\includegraphics{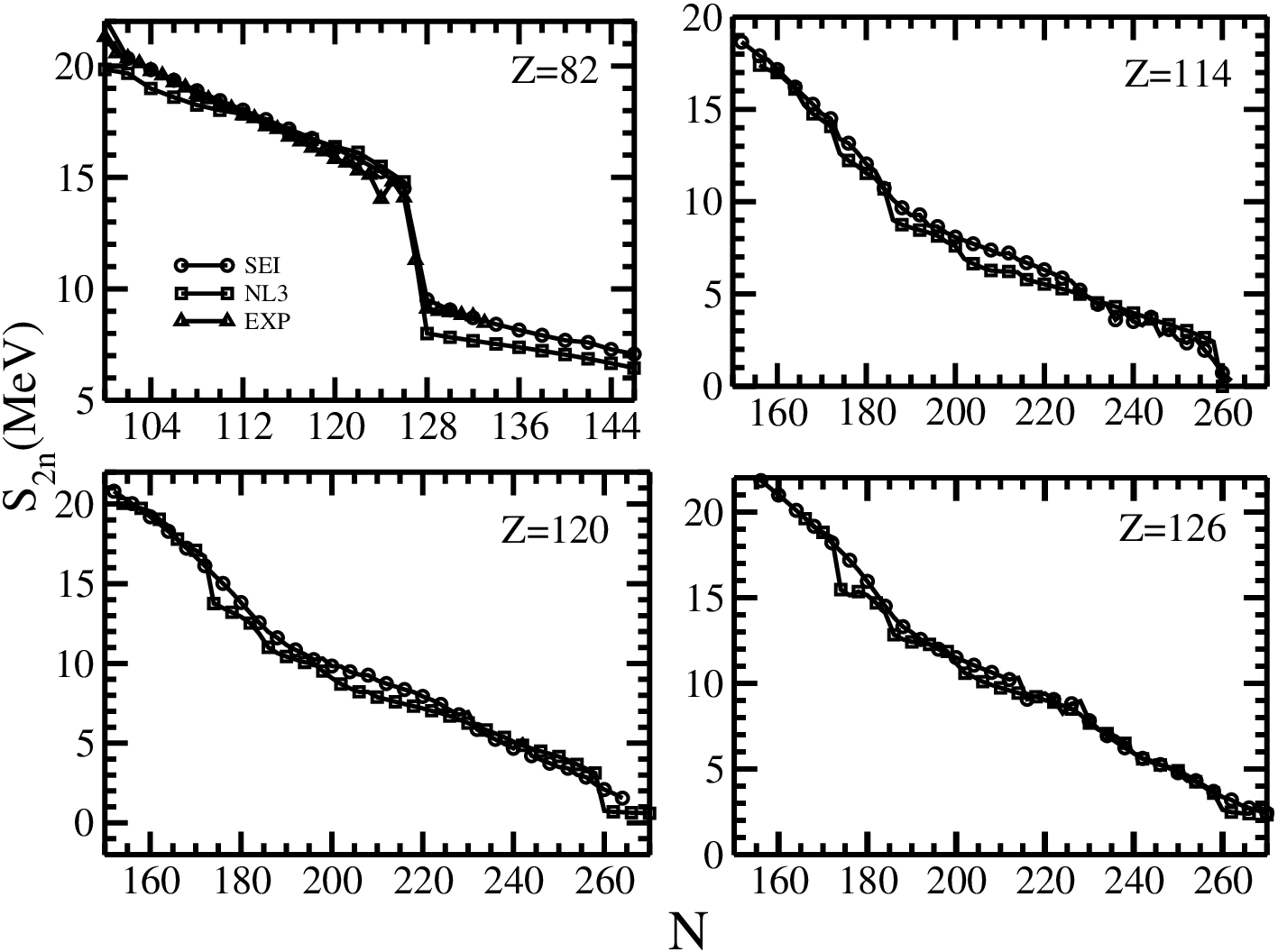}}
\caption{The two neutron separation energy obtained from NL3 and
SEI for $^{208}$Pb, $^{298}$114, $^{304}$120 and $^{310}$126.}
\label{fig:1}
\end{figure}
From the binding energy, we have calculated the two neutron
separation energy using the relation $S_{2n}(N,Z)=BE(N,Z)-BE(N-2,Z)$.
The $S_{2n}$ for all the four isotopic chains are shown in Fig. 2
as a function of neutron number. In case of Pb isotopes, the sudden
decrease of $S_{2n}$  at neutron number {\it N}=126,
is the well known neutron magic number for the largest known {\it Z}=82
magic nucleus. The analysis is extended to the recently predicted
proton magic  numbers like Z=114, 120 and 126, which are currently
under scrutiny for their confirmation.

It is important to mention that, the next proton magic
number beyond {\it Z}=82 would be {\it Z}=126 considering the 
traditional proton and neutron and neutron magic numbers for 
known closed shell nuclei \cite{goldhaber57,myers66}. However, 
several microscopic calculations
\cite{nilsson78,nilsson59,andersson76,meldner67,meldner78a,kohler71}
suggest a shift of this number to 114. One of the cause of the shift is the
Coulomb effect on the spherical single particle levels. The use of shell
correction by V. M. Strutinsky \cite{strutinsky67} to the liquid-drop
calculation of binding energy (BE) opens a more satisfactory
exploration towards the search of double closed nucleus beyond $^{208}$Pb.
Using this approach, {\it Z}=114 is supported to be the proton magic after 82
\cite{nix94,sob94,sob07,smo95}, which was regarded as the magic
number in the superheavy valley \cite{kumar89} with {\it N}=184
as the corresponding neutron magic number. However, the recent relativistic
mean field calculations using various force parameters \cite{bhuyan12},
predict {\it Z}=120 as the next magic number with {\it N}=172/182 as the
neutron closed shell. Contrary to all these predictions, some non relativistic
calculations report {\it Z}=126 as the next magic proton in the superheavy
valley. The microscopic calculations using Skyrme Hartree-Fock
formalism predict {\it N}=182 as the next neutron closed shell after
N=126, which differs by 2 unit from other predictions\cite{bhuyan12}.

\begin{figure}
\vspace{1.0cm}
\resizebox{1.0\textwidth}{!}{\includegraphics{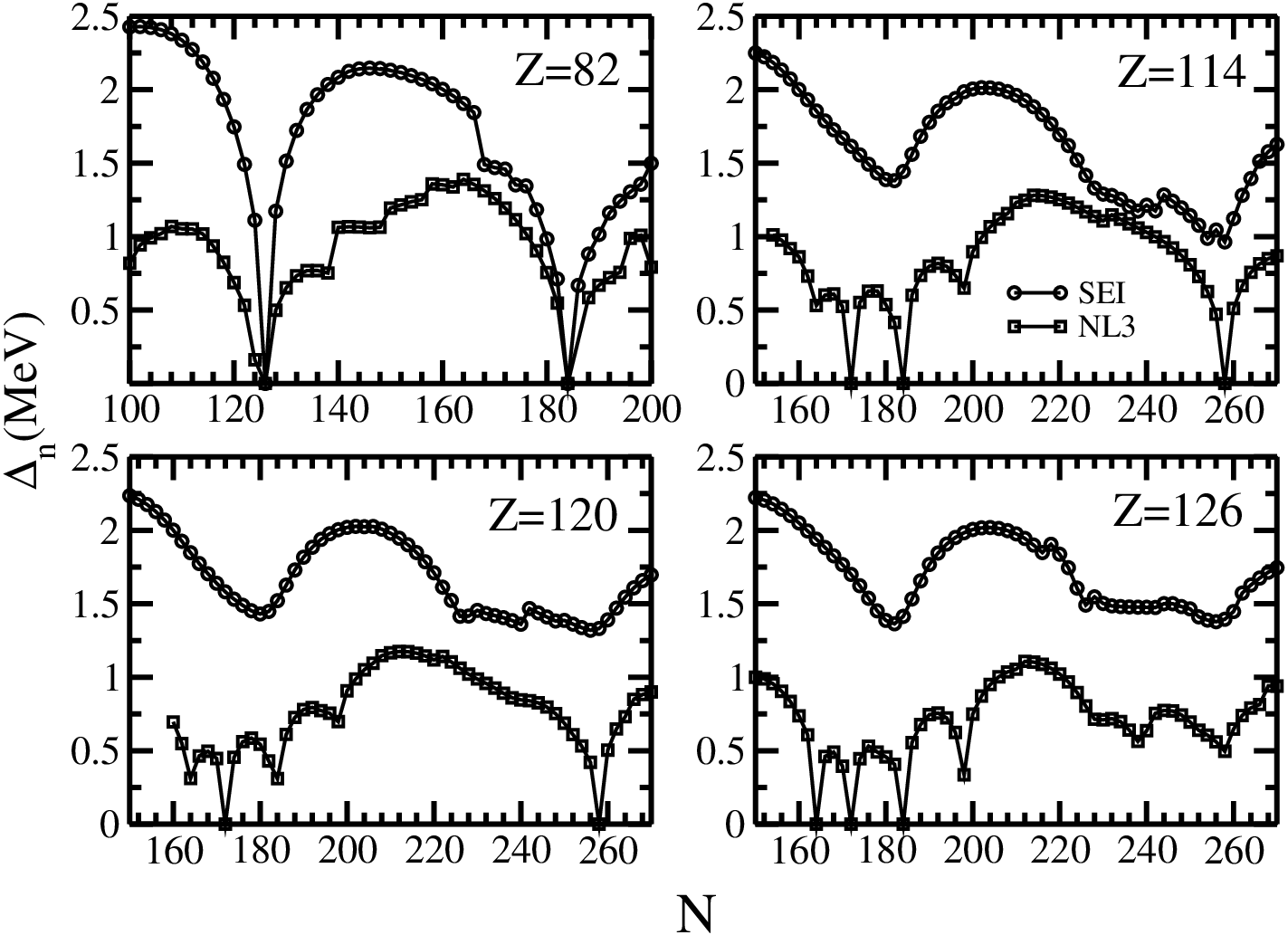}}
\caption{The calculated pairing gap of neutron $\triangle_n$
with SEI for the isotopic series Z=82, 114, 120 and 126 are
compared with the NL3 results.}
\label{fig:1}
\end{figure}

Analyzing  the $S_{2n}$ energy for the isotopic chain
of {\it Z}=82, 114, 120, 126, the sharp fall of $S_{2n}$ at
{\it N}=126 is a clear evidence of magic combination of {\it Z}=82 and
 {\it N}=126. Our newly developed SEI model and previously
existing NL3 follow the  same trend as experiment.
But whenever we analyzed the plots of {\it Z}=114, 120, 126
 find a slight difference in two models (SEI and RMF).
In RMF(NL3), when we go
from one magic neutron number to the next one, the $S_{2n}$ energy suddenly
decreases to a lower value, which reflect in Fig. 2. In SEI, the
$S_{2n}$ energy follows same pattern but the magnitude of decreseness
some how less.

\begin{figure}
\vspace{1.0cm}
\resizebox{1.0\textwidth}{!}{\includegraphics{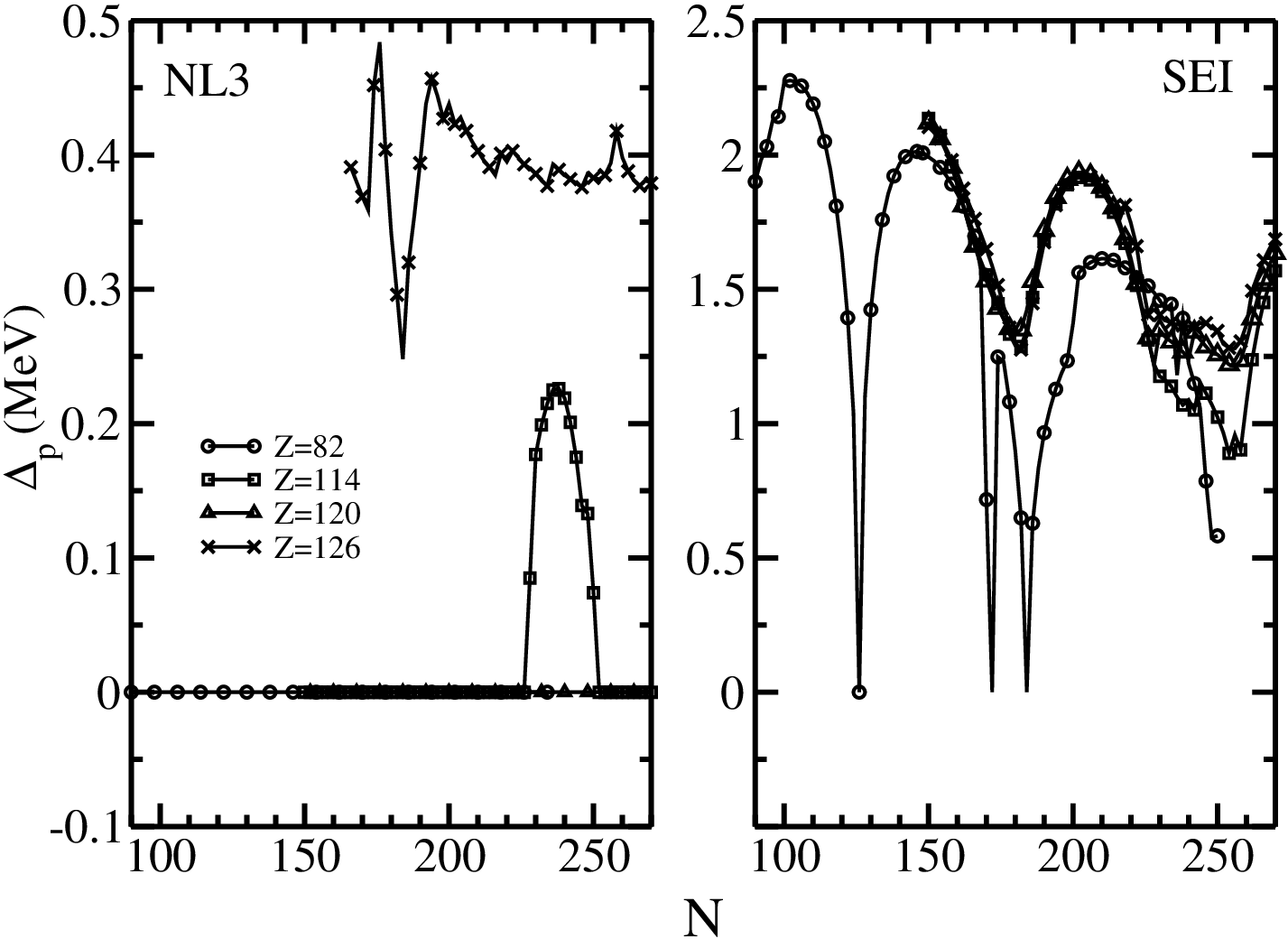}}
\caption{(a)The pairing gap of proton  $\triangle_p$ with SEI
for the isotopic series {\it Z}=82, 114, 120 and 126 compared with
the NL3 results.}
\label{fig:1}
\end{figure}

\subsection{Pairing gaps and pairing energy}

Another important quantity, which helps us to locate the closed shell is
the pairing gaps of proton and neutron in a constant force BCS
calculation. Here, we calculate the pairing gap for the isotopic chain
of {\it Z}=82, 114, 120 and 126 and locate the minimum values of $\triangle_n$
and $\triangle_p$. The results are depicted in Figures 3 and 4 and also
compared with the RMF(NL3) force. It is well known that NL3 force
satisfies this criteria for the location of magicity \cite{sil04,bhuyan12}.
Although, SEI overestimates the paring gaps of  $\triangle_n$,
$\triangle_p$, the trend for both  NL3 and SEI are found to be similar.
Consistence with NL3 results as well as with earlier calculations with
a variety of force parameters, our present SEI reproduces minima at
{\it N}=182/184 and {\it Z}=120 and to some extent at {\it Z}=114.

\begin{figure}
\vspace{1.0cm}
\resizebox{1.0\textwidth}{!}{\includegraphics{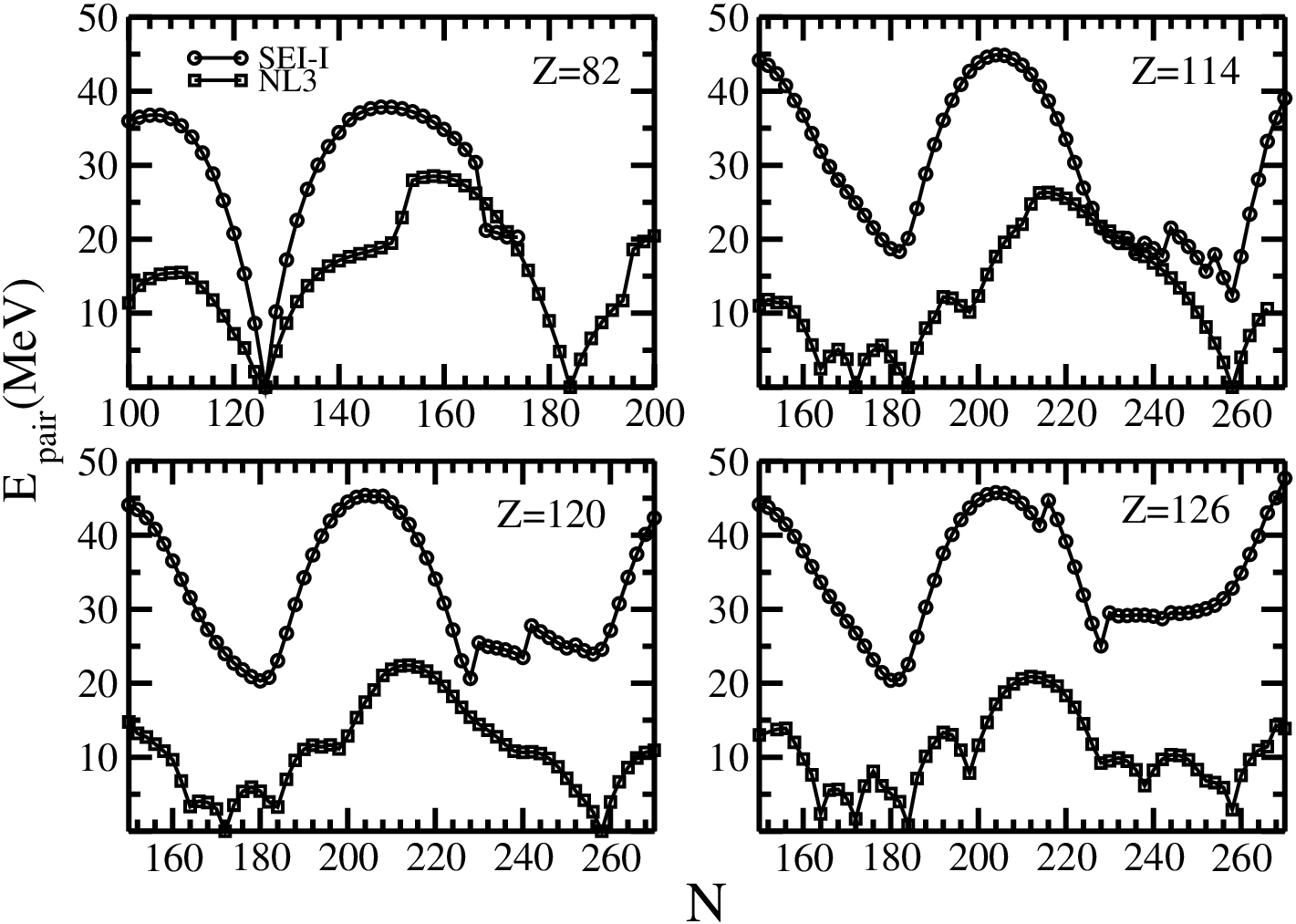}}
\caption{The pairing energy as a function of neutron number
for {\it Z}=82, 114, 120 and 126 with SEI and NL3 forces.}
\label{fig:1}
\end{figure}

To see the trend of pairing energy $E_{pair}$ at the discussed neutron
number {\it N}=184, we plot $E_{pair}$ as a function of neutron number
N in Figure 5. Surprisingly, we get almost zero pairing energy at N=126
for {\it Z}=82 isotopic case. The formalism is extended to {\it Z}=114,
120 and 126 cases. We find minimum or zero $E_{pair}$ at {\it N}=182/184
confirming the earlier predictions of this neutron magic number at {\it N}
=182/184 \cite{bhuyan12}.Qualitatively, the SEI interaction follows the
trend of RMF(NL3) as shown in Fig 5, but fails when we have a quantitative
estimation. For example, the $\triangle_n$ or $E_{pair}$ at {\it N}=182/184
is minimum but has a finite value unlike to the NL3 prediction. As a matter
of fact, the validity of pairing scheme to this region of nuclei may not be
100 {\% } applicable. The importance of pairing is needed to keep the value
of $\triangle _n$ and $\triangle_p$ zero at the appropriate magic number.

\begin{figure}
\vspace{1.0cm}
\resizebox{1.0\textwidth}{!}{\includegraphics{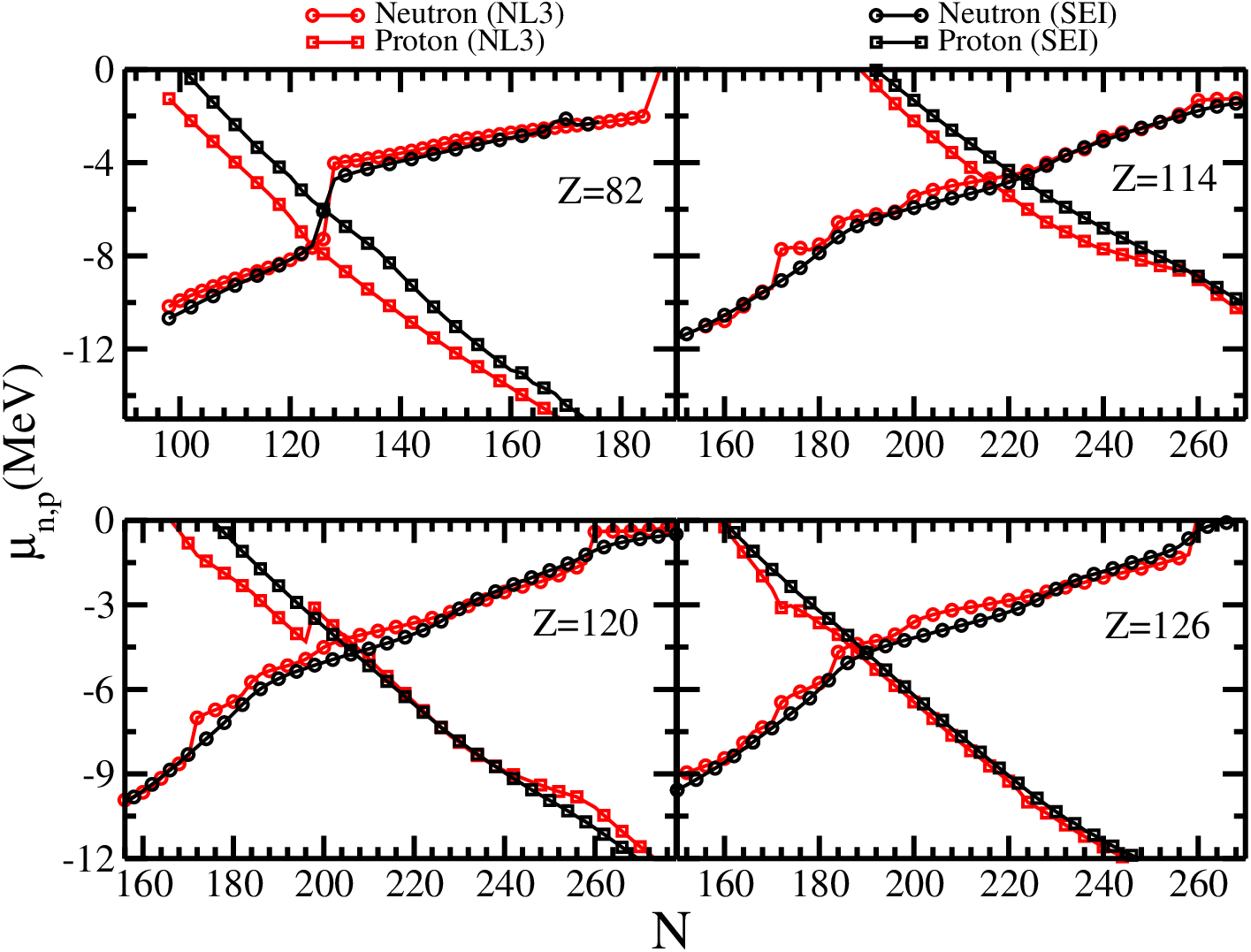}}
\caption{Systematics of chemical energy $\mu_n$ and $\mu_p$
as a function of neutron number for {\it Z}=82, 114, 120 and
126 with SEI and NL3 sets.}
\label{fig:1}
\end{figure}

\subsection{Chemical energy and stability}
It is to be noted that one can find similar information about 
the stability of a nucleus either from the chemical potential 
or the nucleon separation energy. However, the neutron or proton 
separation energies are obtained from the binding energy, whereas 
the chemical potential (both for proton and neutron) calculated 
self-consistently while solving the field equations. To see the 
consistency between these two observables ($S_{2q}$ or $\mu_q$; 
$q$=$n,p$), we have analyzed these quantities separately in the 
present paper. For a bound nucleus, both the chemical potentials 
of protons $\mu_p$ and neutrons $\mu_n$  must be negative. To 
realize the relative stability
from chemical point of view, we have plotted $\mu_p$ and $\mu_n$
with neutron number in Fig. 6. The results are also compared with the
$\mu-$value of NL3 set. In both the cases, we find similar chemical
potential. In some previous papers it was suggested that we can take
{\it N}=172 as magic number for neutron. But our SEI model show that the
combination {\it Z}=120 and {\it N}=172 is strictly not allowed. Because
in this case $\mu_p=0.69$ MeV, which gives proton instability. However
NL3 result shows this combination is a loosely bound system having
$\mu_p=-1.240$ MeV and $\mu_n=-7.007$ MeV. Although the BE/A curve show
a local maximum at {\it Z}=114 and N=172 in SEI model, we can not take
this as a stable system because of $\triangle_n$ and $\triangle_p$ value, which
does not shows any signature of stability. The SEI model gives a clear picture
that the isotope $^{302}120$ be, a suitable combination for the next double
closed nucleus. One can justify it by analysis of BE/A data of $^{302}120$.
For example $BE/A=7.007$ MeV which create a local maxima in its neighbor hood for
$^{302}120$. In the same time, the optimum negative value of chemical potential
energies of $\mu_n$and $\mu_p$ gives a sign of maximum stability. A similar
analysis of numerical data for $\mu_p$ of isotopes of {\it Z}=126 shows that
there is no reason of taking {\it Z}=126 and {\it N}=182/184 as a stable
combination. This is because of the positive value of $\mu_p$ (1.36) MeV.

\begin{figure}
\vspace{1.0cm}
\resizebox{1.0\textwidth}{!}{\includegraphics{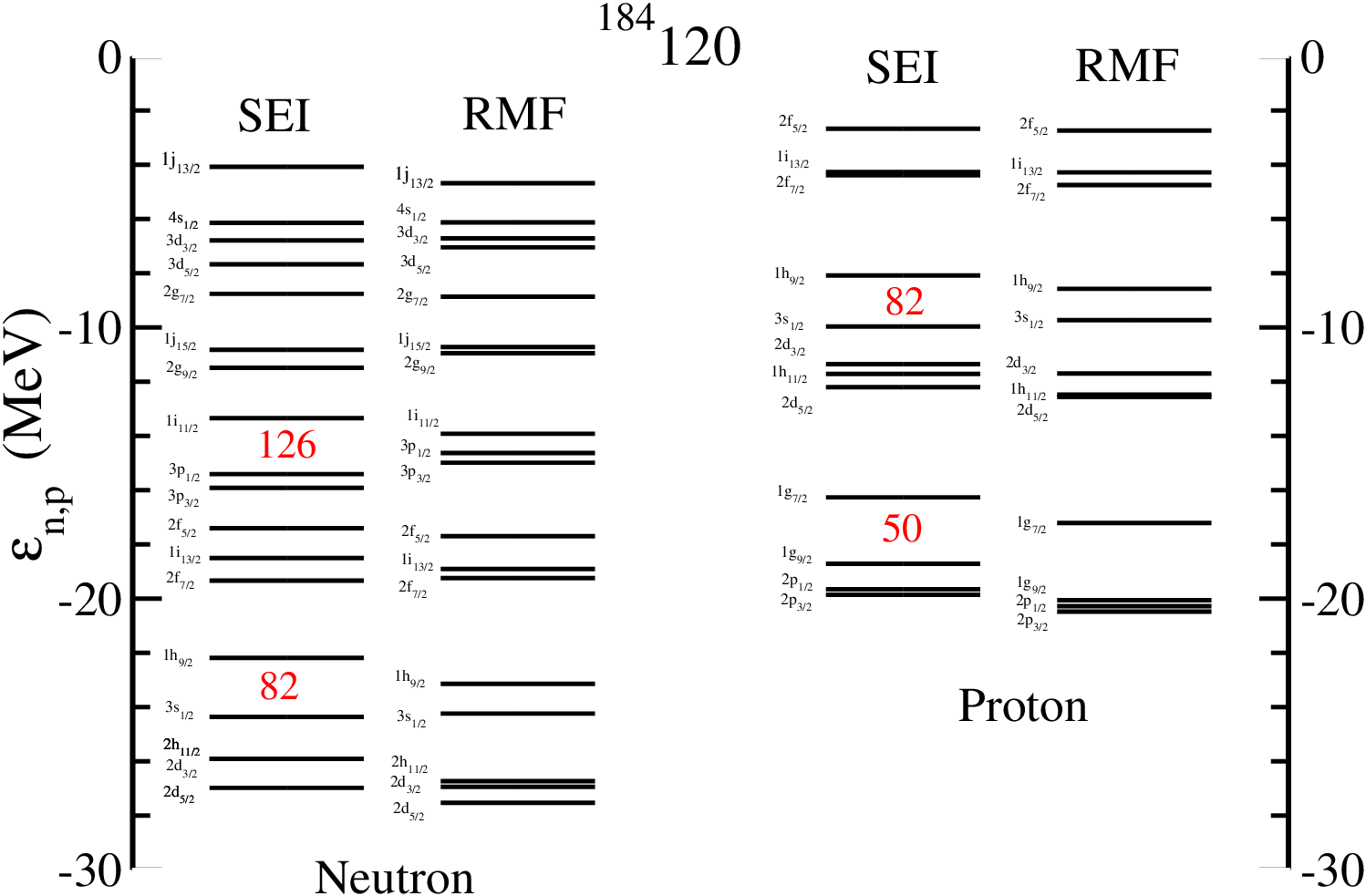}}
\caption{ The single particle energy levels for $^{304}$120 with
NL3 and SEI parametrization.}
\label{fig:1}
\end{figure}

\subsection{Single particle energy}
The single particle energies for $^{304}$120 with NL3 and SEI
for proton and neutron are shown in Fig 7. The single particle
solutions are obtained without including the pairing correlation
into account to intact the degeneracy of the levels. The calculation
of single particle energies of SEI with pairing shows that the degeneracy
of the energy levels are not invariant. The basic cause of this discrepancy
is the over estimation of our pairing strength in SEI model which may
be an interesting analysis for pairing in future. The filling up single
particle energy levels for neutrons in SEI with pairing is different from
that of without pairing. The energy levels without pairing are given by
[178] $({{3d}_{3/2}})^4$, $({{4s}_{1/2}})^2$
while the same with pairing are [178] $({{3d}_{3/2}})^3$, $({{4s}_{1/2}})^1$,
$({{1j}_{11/2}})^2$. That means an empty orbital is created at 4$s_{1/2}$ and
occupied in 1$j_{11/2}$. We have also analyzed the single particle levels for
$^{302}$120, which is not given in the figure. From the anatomy of $\epsilon_n$
and $\epsilon_p$, we find large gaps at neutron number {\it N}=184 and proton
number {\it Z}=120.  The value of neutron gap at {\it N}=184 is 1.949 $MeV$
and that of proton is 1.275 $MeV$ for the last occupied and first unoccupied
nucleon. On the other hand the neutron and proton gap for $^{302}$120 are
respectively  $\sim 0.6$ and $\sim 1.663$$ MeV$. The above data say the
energy gaps for the neutron and proton in $^{304}$120 are greater than
the  gap in $^{302}$120. This give us an indication to take the combination
{\it N}=184 and {\it  Z}=120 as the next magic nucleus. From the  analysis
of single particle energy level of $^{304}$120 with NL3 parameter set one
can see the neutron and proton gaps are $1.4503$ MeV and $2.1781$ MeV
respectively for the last occupied and first unoccupied nucleon.
 The RMF(NL3) and SEI data are comparable with each other.

\section{Summary and Conclusions}
In summery, we have calculated binding energy $S_{2n}$ energy, single
particle levels, pairing gaps and chemical potential, in the isotopic chain of {\it Z}=82,
114, 120 and 126. All our calculations are done in the frame work of SEI
interaction. We have compared our results with standard RMF(NL3) interaction.
Over all discussion and analysis of all possible  evidences of shell closure
 property show that, one can take {\it Z}=120 and {\it N}=182 as the next magic combination beyond $Z=82$ and $N=126$, which is different from Skyrme, Gogny, RMF(NL3) by two unit. However on the basis of single particle energy levels, the preferred gap is at
{\it N}=184 which is consistent with these (Skyrme, Gogny and RMF) force
parameters. This happens due to the overestimation of pairing strength. As
we only use SEI interaction to predict the magic nuclei, so it is out of
scope of our status to change its original pairing strength. We can just make
a comments to this observation and left for further study.

\section{Acknowledgments}
We thanks Ms. Shikha for careful reading of the manuscript.

\end{document}